# Hollow-core fibers with ultralow loss in the ultraviolet range and sub-thermodynamic equilibrium surface-roughness

**J. H. Osório[1], F. Amrani[1, 2], F. Delahaye[1, 2], A. Dhaybi[1], K. Vasko[1], G. Tessier[3], F. Giovanardi[4], L. Vincetti[4], B. Debord[1, 2], F. Gérôme[1, 2], F. Benabid[1, 2]**

[1]*GPPMM Group, XLIM Institute, CNRS UMR 7252, University of Limoges, Limoges, France.*
[2]*GLOphotonics, 123 Avenue Albert Thomas, Limoges, France.*
[3]*Institut de la Vision, CNRS UMR 7210, INSERM, Sorbonne University, Paris, France.*
[4]*Department of Engineering 'Enzo Ferrari', University of Modena and Reggio Emilia, Modena, Italy*

*f.benabid@xlim.fr*

**Abstract:** We report on hollow-core fibers displaying rms core roughness under the surface capillary waves thermodynamic limit and record loss in the short-wavelength range (50.0 dB/km at 290 nm, 9.7 dB/km at 369 nm, 5.0 dB/km at 480 nm, 0.9 dB/km at 558 nm, 1.8 dB/km at 719 nm). © 2021 The Author(s)

Among the novel frontiers in photonics are ultraviolet (UV) photonics and next-generation optical fibers. The former holds a prominent position within technological and scientific areas, as it includes activities that entail or require the generation and control of light at wavelengths between 400 nm and 100 nm. Such activities encompass materials synthesis, micro- and nano-processing, precision spectroscopy, and light transport. The latter saw a huge progress, exemplified by the latest achievements in inhibited-coupling (IC) hollow-core photonic crystal fibers (HCPCFs) technology. Still, although the recent results on these fibers' performances are remarkable, further advances should rely on the improvement of the HCPCFs core surface quality as it imposes a fundamental hindrance to loss decreasing due to scattering processes. Differently from the IC HCPCFs guiding in the infrared (IR), whose loss remains limited by the fiber design (*i.e.*, confinement loss, CL), the current state-of-the-art in HCPCFs with guidance at short wavelengths (λ < 800-1000 nm) is constrained by the surface scattering loss (SSL) [1]. Thus, although the cladding design optimization efforts have entailed significant lessening of the loss for fibers guiding in the IR, reducing it in the visible and UV ranges lingers as a more challenging task owing to SSL limitation.

SSL is set by the roughness of the core surface, which stems from frozen-in thermal surface capillary waves (SCW) that are forged during the fiber draw. SSL takes the form of Eq. (1), where $F$ is the core mode and core contour overlap, $\lambda$ is the wavelength, and $\lambda_0$ is a calibration constant, and scales with the square of surface roughness root-mean-square (rms) height, *i.e.*, $\alpha_{SSL} \propto h_{rms}^2$ [2]. The multiplying factor $\eta$ in Eq. (1) relates to the surface quality and is proportional to $h_{rms}^2$. The thermodynamic equilibrium surface roughness (TESR), $h_{TESR}$, amounts to $\sqrt{k_B T_G/\gamma}$ ~ 0.4 nm for silica ($k_B$: Boltzmann constant, $T_G$: glass transition temperature; $\gamma$: interfacial tension).

$$\alpha_{SSL} = \eta \times F \times \left(\frac{\lambda}{\lambda_0}\right)^{-3} \tag{1}$$

Thus, diminishing the SSL levels must rely on manipulating the surface roughness. In this context, recent results have shown that the roughness along the fiber drawing direction can be lowered by controlling the drawing stress [3]. Furthermore, SSL reduction enables the advent of optical fiber with unprecedentedly low transmission-loss and solarization-resistance, which, in turn, gives UV-photonics a strong leverage to further industrial and scientific applications.

Here, we report on the fabrication of HCPCFs with improved core surface quality and record low loss in the short-wavelength range. To accomplish it, we revisited the usual HCPCF fabrication methods and obtained a rms roughness reduction from ~0.4 nm in fibers drawn with standard fabrication methods to 0.15 nm in fibers fabricated by using an innovative technique. The new fibers have, therefore, sub-TESR levels. The reduction in the core roughness allowed attaining fibers with ultralow loss: 50.0 dB/km at 290 nm, 9.7 dB/km at 369 nm, 5.0 dB/km at 480 nm, 0.9 dB/km at 558 nm, 1.8 dB/km at 719 nm. We envisage that our investigation will confer a new panorama for fibers working in short-wavelengths and a new path for attaining ultra-smooth surfaces for next-generation applications in photonics.

Fig. 1 shows the microstructures of the fibers we report herein (F#1 and F#2) and a zoom in the lowest loss regions. The fibers are tubular amorphous lattice cladding HCPCF formed by 8 non-touching tubes. Table 1 lists the core diameter, $D_{core}$, the thickness and diameter of the cladding tubes $t$ and $D_{tubes}$, and the spacing between the lattice tubes, $g$, of the fibers. F#1 exhibits minimum loss values of 50.0 dB/km at 290 nm, 9.7 dB/km at 369 nm, and 5.0 dB/km at 480 nm. The minimum attenuation values for F#2 are 0.9 dB/km at 558 nm and 1.8 dB/km at 719 nm. Fig. 2 displays the measured loss of F#1 and F#2, the silica Rayleigh scattering limit (SRSL) trend, and a typical output near field profile. Noteworthily, the values reported herein are new record low loss figures for fibers working in the short-wavelength range [4-8] and lie below the Rayleigh scattering limit of bulk silica (in red).

Table 1. F#1 and F#2 geometrical parameters ($D_{core}$: core diameter; $t$: thickness of the cladding tubes; $D_{tubes}$: diameter of the cladding tubes; $g$: spacing between the lattice tubes).

|  | F#1 | F#2 |
| --- | --- | --- |
| $D_{core}$ (µm) | 27 | 42 |
| $t$ (µm) | 0.6 | 0.9 |
| $g$ (µm) | 2.1-4.7 | 3.6-5.2 |
| $D_{tubes}$ (µm) | 11 | 18 |

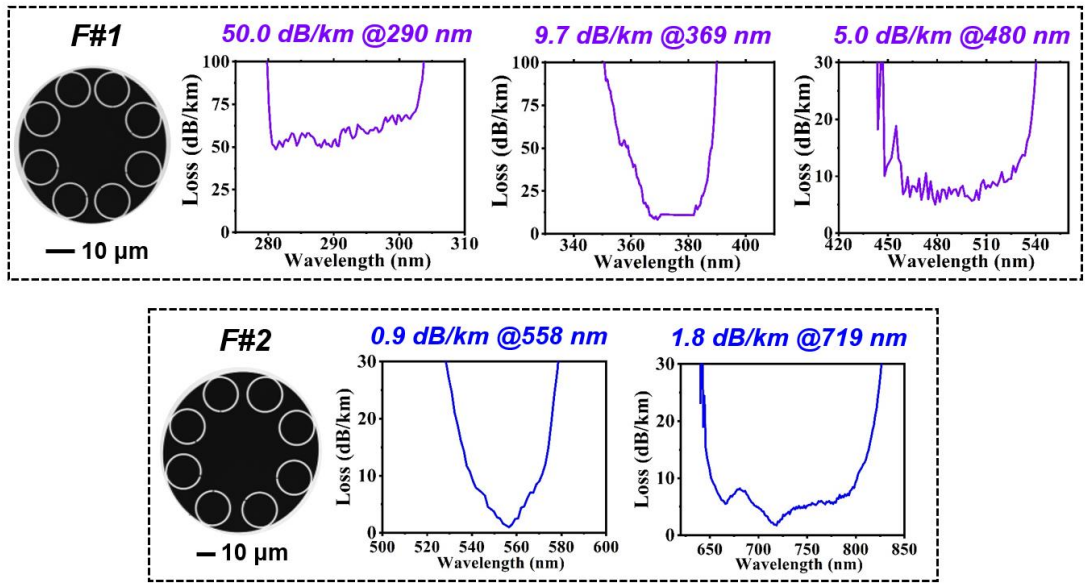

Fig. 1. F#1 and F#2 cross-sections and measured loss. Minimum loss values are 50.0 dB/km at 290 nm, 9.7 dB/km at 369 nm, 5.0 dB/km at 480 nm, 0.9 dB/km at 558 nm, 1.8 dB/km at 719 nm.

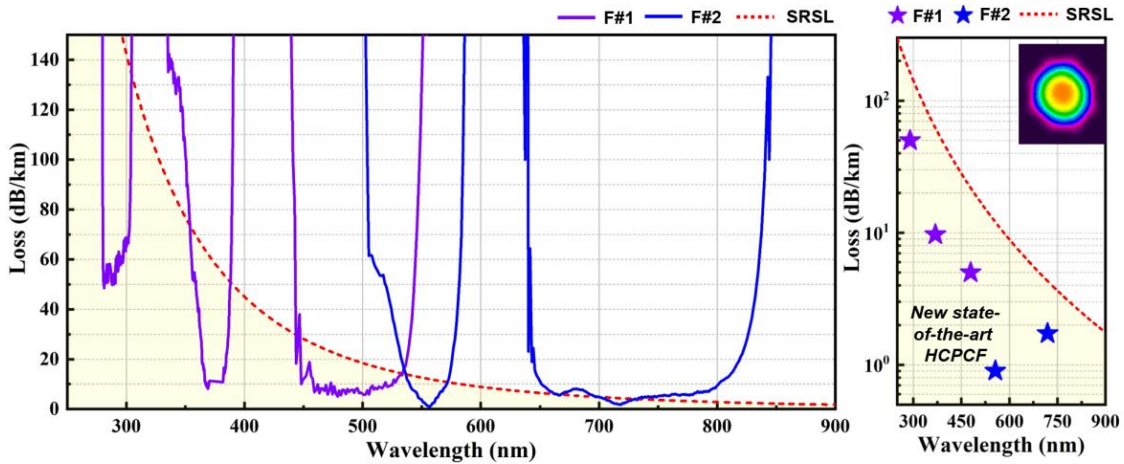

Fig. 2. F#1 and F#2 measured loss together with the silica Rayleigh scattering limit (SRSL) trend and a typical output near field profile.

The remarkable ultralow loss reported above could be attained thanks to an amelioration of the core surface quality, which was accomplished by controlling the shear stress acting onto the cladding microstructure during the draw [2]. To study the effectiveness of the new fabrication approach, we assessed, by using a picometer resolution profilometer [9], the core roughness in two groups of tubular amorphous lattice cladding HCPCFs, G#1 and G#2. G#1 fibers have been drawn by using the standard procedure and G#2 fibers have been drawn by using our new methods.

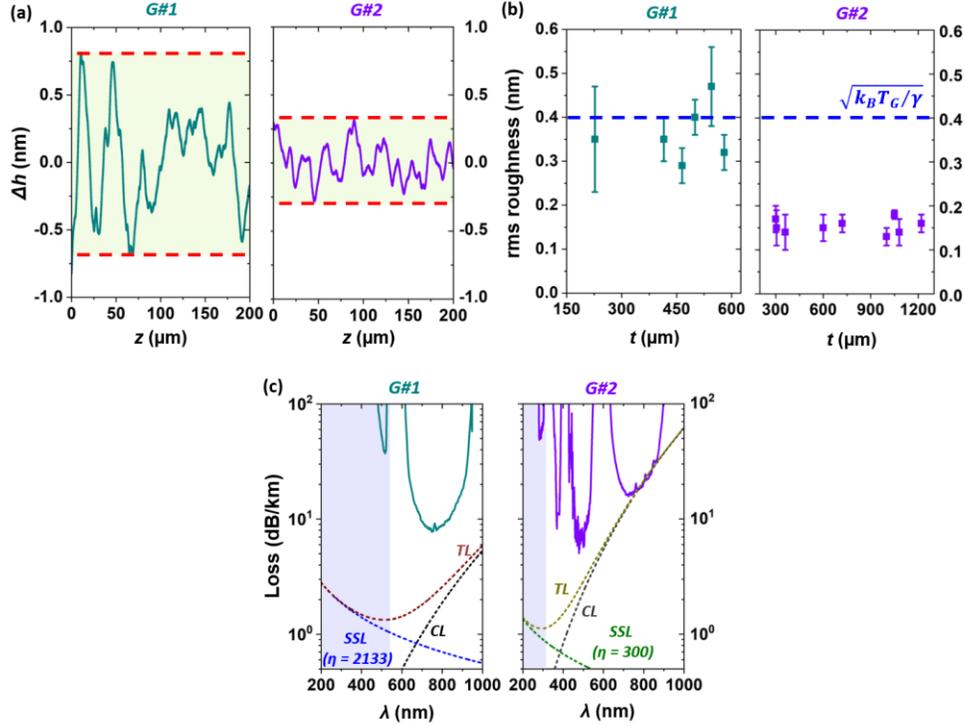

Fig. 3. (a) Typical core surface roughness profiles (Δh: roughness height variation) and (b) measured rms roughness. (c) Typical loss spectra for fibers in G#1 and G#2; Confinement loss (CL), surface scattering loss (SSL) and total loss (TL) trends.

Fig. 3a exhibits typical surface profiles along the HCPCF axis of G#1 and G#2 fibers. Fig. 3a apprises that the peak-to-peak roughness of G#1 fibers has values around 1.5 nm while, for G#2 fibers, values around 0.5 nm. Fig. 3b displays the rms roughness of G#1 and G#2 fibers with different cladding tube thicknesses estimated from several surface roughness measurements. The rms roughness of G#1 fibers fluctuates around 0.4 nm. G#1 fibers have, thus, TESR. In turn, G#2 fibers have rms roughness around 0.15 nm, which is considerably lower than the TESR limit set by the SCW scenario in amorphous silica (Fig. 3b, blue line).

Fig. 3c shows the typical loss for G#1 and G#2 fibers with similar tubes thickness. The distinction between the loss trends is patent. The loss in the reported G#1 fiber ($D_{core}$ = 41 μm) presents a considerable loss increase for λ < 750 nm due to SSL influence. In the reported G#2 fiber ($D_{core}$ = 27 μm), a decreasing loss trend is observed even for λ < 750 nm. Fig. 3c also shows the CL, SSL, and total loss (TL = CL + SSL) trends calculated by considering the scaling laws in [10]. Here, we used η = 2133, corresponding to a fiber surface with TESR (in blue), and η = 300, corresponding to a surface-roughness rms that is 2.7 lower than the TESR rms (in green). The η values have been estimated by adjusting its value to the loss of fibers displaying TESR and sub-TESR [2, 3, 11]. It is seen that the TL turning point happens at shorter wavelengths for smaller SSL. The difference between the loss behaviors of G#1 and G#2 fibers is, hence, a signature of SSL reduction in the new fibers.

In summary, we developed HCPCFs exhibiting ultralow loss at short wavelengths by improving the fiber cores' surface quality and by attaining sub-TESR levels. We envision that our research will provide a new scenery for the next HCPCFs for the visible and ultraviolet ranges and a new avenue for attaining next-generation photonic devices with ultralow surface roughness levels.

**Acknowledgements**

This research was funded through PIA program (grant 4F) and la region de la Nouvelle Aquitaine.